# A Generalized Publication Bias Model

Peter H. Schonemann[1]      Jeffrey D. Scargle[2]

[1]Department of Psychology      [2]Space Science Division

Purdue University      NASA Ames Research Center

West Lafayette IN 47906 USA      Moffett Field CA 94035-100 USA

phs@psych.purdue.edu      jeffrey@cosmic.arc.nasa.gov

KEY WORDS: publication bias, meta-analysis, file-drawer hypothesis, fail-safe number.

Scargle (2000) has discussed Rosenthal&Rubin's (1978) "fail-safe number" (FSN) method for estimating the number of unpublished studies in meta-analysis. He concluded that this FSN cannot possibly be correct because a central assumption the authors used conflicts with the very definition of publication bias. While this point has been made by others before (Elsahoff, 1978; Darlington, 1980; Thomas, 1985, Iyengar & Greenhouse, 1988), Scargle showed, by way of a simple 2-parameter model, how far off Rosenthal & Rubin's estimate can be in practice. However, his results relied on the assumption that the decision variable is normally distributed with zero mean. In this case the ratio of unpublished to published papers is large only in a tiny region of the parameter plane.

Building on these results, we now show that

(1) Replacing densities with probability masses greatly simplifies Scargle's derivations and permits an explicit statement of the relation between the probability $\alpha$ of Type I errors and the step-size $\beta$;

(2) This result does not require any distribution assumptions;

(3) The distinction between 1-sided and 2-sided rejection regions becomes immaterial;

(4) This distribution-free approach leads to an immediate generalization to partitions involving more than two intervals, and thus covers more general selection functions.

# 1. Introduction: Historical Context

In the late 70's, Rosenthal & Rubin (1978) and Rosenthal (1979) proposed a new method for coping with the nagging "file-drawer problem" of meta-analysis. In essence, meta-analysis is a quantitative method for aggregating statistical results of a number of similar studies on a particular topic into a hopefully more conclusive larger study, relying on methods proposed by Wallis (1942), Fisher (1948), and others. These procedures are statistically sound as long as the necessary assumptions - most importantly the fairness of the samples - are met at least approximately. Problems arise when they are not met, as in meta-analysis, a supposedly more objective method for evaluating accumulated research.

Mahoney (1977) has defined "confirmatory bias" as "the tendency to emphasize and believe experiences which support one's views or discredit those which do not" (p. 161). This type of bias has been repeatedly verified and seems to be pervasive and quite robust. After asking 75 journal reviewers "to referee manuscripts which described identical experimental procedures but which reported positive, negative, mixed or no results", he not only found poor inter-rater agreement, but also, more to the point here, that "reviewers were strongly biased against manuscripts which reported results contrary to their theoretical perspective" (*loc. cit.*)

In an effort to cope with this challenge, Rosenthal & Rubin (1978) proposed a so-called "fail-safe number" (FSN) approach intended to estimate *post hoc* the number of unpublished papers that languish in file-drawers because they had been rejected as a result of confirmatory bias. The implied claim was that a positive effect in the published sample could only be due to publication bias if there were at least this number of unpublished papers. Typically, this FSN turned out to be very large. For example, in the first paper published on this topic in *Behavioral and Brain Sciences,* (BBS), Rosenthal & Rubin (1978) reported a FSN of 65,122 for 345 published studies. The authors concluded that "It certainly seems unlikely that there are file drawers crammed with the unpublished results of over 65,000 studies of interpersonal expectations" (p. 381).

One of the commentators, Elsahoff (1978), presented a simple counterexample in the same issue that suggested the FSN logic had to be flawed, since Rosenthal & Rubin's FSN appeared to be far too large. One year later, Rosenthal (1979) restated this FSN argument in the more widely read *Psychological Bulletin*. This paper spawned a veritable avalanche of meta-analytic literature reviews, which continues unabated to this day, and which is predicated on the mistaken belief that Rosenthal & Rubin had banished the confirmatory publication bias problem once and for all.

Barely one year after Rosenthal's paper had appeared, Darlington (1980) submitted a manuscript to the *Bulletin* challenging Rosenthal & Rubin's reasoning with the following simple counterargument: "… imagine that all the test null hypotheses in a certain area are true, and that the only results published are the 5% of studies which achieve significance by chance. Suppose that 345 studies were published this way (The number 345 is used to facilitate later comparison with the Rosenthal-Rubin example.) Then we are imagining that the total number of studies performed was T = 20x345 or 6900" (loc. cit. p.3) - not 65,467, as Rosenthal & Rubin had reported.

Darlington also furnished a plausible explanation for the discrepancy between his and Rosenthal & Rubin's findings: Rosenthal & Rubin had simply made a mistake: "The specific statistical assumption Rosenthal adopted is inconsistent with the problem he purported to solve … Rosenthal's formulation assumed that the mean Z of the unpublished studies would be zero. But if all the most significant studies were published, and if the mean Z of all studies (published and unpublished) were zero, then the mean of the unpublished studies would be negative" (loc. cit.).  Although at least one referee had encouraged the editor to publish Darlington's paper, the *Bulletin* editor ended up rejecting it on the advice of other reviewers.

Unaware of Darlington's precedent, a second author, Thomas (1985), submitted a very similar critique to the *Psychological Bulletin* " … the distribution of Z scores, conditioned on their appearing in the literature, cannot be standard normal in distribution [but rather a] truncated distribution of the original test statistic distribution." (Abstract, p.2). Thomas, too, was unable to convince the new *Bulletin* editor of the validity of this simple argument. After encouraging him to revise his manuscript several times, the *Bulletin* ended up rejecting it after almost five years of protracted review.

A whole decade elapsed since Rosenthal & Rubin's introduction of their FSN method before Iyengar & Greenhouse (1988) finally managed to straighten out the public record. Starting out with fulsome praise for "Rosenthal's fail-safe sample size approach [as] an elegant formulation," they eventually noted  "several drawbacks that limit its usefulness" (p.115). One of them was that Rosenthal (1979)  "assumes that the mean Z value for the unpublished studies is zero … Now, if there were publication bias in favor of studies with statistically significant findings, the Z values would not be a sample from a standard normal distribution" (p.111). This was precisely the point that both Darlington (1980) and

Thomas (1985) had vainly tried to register ten years earlier.                                 6

As an alternative to the FSN approach, Iyengar & Greenhouse proposed fitting a selection function accommodate the bias. As they acknowledge, such a procedure vitiates the main selling point of Rosenthal and Rubin's FSN method, its computational simplicity: "The maximum likelihood approach based on selection models does involve much more computation" (Iyengar & Greenhouse, 1988, p.115).

Scargle (2000) expanded on their model. Specifically, he investigated the assumption that publication bias is a function only of a single variable, the reported z-score summarizing the study. The quantitative model he adopted was a 2-parameter step-function for the publication probability:

(1)      $S(z) = \text{Prob}(\text{publication}) = \beta$, if $z < z_0$,   $S(z) = 1$,   if $z \geq z_0$.

The publication probability as a function of z is found by multiplying the selection function S(z) discussed above by the probability distribution of the reported z values themselves, denoted here g(z). For example, the null hypothesis might be that g(z) is a zero-mean normal distribution.

On defining

(2)           $p :=$ probability that a paper will be published, the ratio

(3)                         $(1-p)/p := r$

plays a central role in Scargle's argument. He found that r is large only for a tiny region of the (z, β) plane (Fig. 3 of Scargle, 2000, with different notation. See also Table 1). This fact undermines the FSN rationale of Rosenthal & Rubin.

========================
Table 1 and Figure 2 about here

To fix terms, let us call r "large" if it exceeds 50, and "very large" if it exceeds 100. Table 1 shows that r is never large for any of the conventional significance levels ($\alpha$ = .01, .05, .10), while Rosenthal & Rubin's r = 189 is very large.

We now generalize Scargle's earlier results in two directions: We show (1) that the main result can be obtained without any distribution assumption, and (2) that such a distribution-free approach easily extends to more general selection functions S(z), thereby removing possible concerns that the original 2-parameter family may be too restrictive.

## 2. Main Result

For added clarity, we restate the main features of Scargle's step-function selection model in Fig. 1. Beyond a certain cut-off $z_0$, all studies are published with probability 1, below this cut-off, they are published with constant probability beta. Geometrically, $\beta$ is the step-size of the selection function S(z) in eq. (1). Assuming the density g(z) is normal, Scargle studied the contours of r in a ($\beta$, $z_0$) coordinate system (Fig. 3 in Scargle, 2000). We now show that they can be studied without any assumptions about the shape of g(z).

==============
Figure 1 about here
==============

To begin with, note that for any given density g(z), $\alpha$ serves just as well as $z_0$ to represent the location of the step in S(z). We therefore may transform to a more convenient $(1-\alpha, 1-\beta)$ coordinate system. For the sake of simplicity, we treat the 1-tailed case where $\alpha$ is the probability mass under g(z) to the right of $z_0$, and $\beta$ is the constant probability that a paper with $z < z_0$ is published (Fig. 1). Our conclusions apply equally to the two-sided case, where $\alpha$ includes both the upper and lower tails of g(z). This forestalls unnecessary disputes about the 1-tailed versus the 2-tailed case that concerned Rosenthal & Rubin (1988, p.120) versus Iyengar & Greenhouse (1988, p. 134).

The total probability that a paper is rejected is then, as a fraction of a fraction,

(4) $\qquad 1 - p = (1-\alpha)(1-\beta),$

and the complementary probability p of publication is

(5) $\qquad p = 1 - (1-\alpha)(1-\beta).$

Eq. (17) in Scargle (2000) is a special case of eq. (5) with g a normal distribution. Here, however, we have made no distribution assumptions at all. To obtain an expression for r in eq.(3), we divide the identity

(6) $\qquad (1-p) = (1-p)/[(1-p) + p]$

by p, arriving at

(7) $\qquad 1 - p = r/(r+1).$

Therefore, for fixed $r = r_0$, one obtains from eq.(4)

(8) $\qquad 1 - p = (1-\beta)(1-\alpha) = r_0/(r_0+1).$

This equation is basic for all further derivations. The contours $r = r_0$ describe a family of rectangular hyperbolae in a $(1-\beta, 1-\alpha)$ coordinate system. Since the range of both variables is restricted to the unit square, they resemble a family of straight lines with

Publication Bias and File-drawer Problem

slope -1 (cf. Figure 3). Figure 2 shows a plot of (ln) r as a function of the two parameters 9 $\alpha$ and $\beta$. Nowhere in the parameter space covered in Table 1 is r "very large".

========================
Figure 3 and Figure 3 about here
========================

## 3. Extension to more general selection functions S(z)

From eq. (5) one further finds that

(9) $\qquad p = \beta(1 - \alpha) + 1\, \alpha,$

which is just

(10) $\quad p = \text{Prob}(\text{publish} \mid z \in Q_1)\, \text{Prob}(z \in Q_1) + \text{Prob}(\text{publish} \mid z \in Q_2)\, \text{Prob}(z \in Q_2),$

where $Q_1$, $Q_2$ partition the real line into two intervals separated by $z_0$, and Prob( publish | $z \in Q_k$ ) denotes the conditional probability that a paper is published if it falls into the partition $Q_k$ under the otherwise unspecified density g.

Eq. (10) suggests an immediate generalization to partitions $Q_k$ with more than two intervals and thus to more general selection functions S(z). Let

(11) $\qquad A := [A_1,\ A_2\ \ldots\ A_m], \qquad$ with $A_k := \text{Prob}(z \in Q_k)$

be a piecewise constant representation, or approximation, of the density function g, and

(12) $\qquad B := [B_1, B_2, \ldots, B_m], \qquad$ with $B_k := \text{Prob}(\text{published} \mid z \in Q_k)$

be the associated conditional probabilities. Then p is simply the sum

(13) $\qquad p = \text{Prob}(\text{published}) = \Sigma_k A_k B_k \qquad k = 1, \ldots, m.$

As an example, consider

(14)    $A = [.6, .3, .1]$    and    $B = [.2, .4, .9]$.

Eq. (14) defines a J-shaped step-function, where *e.g.* $B_3 = .9$ reflects the fact that perhaps not even papers reporting highly significant results are published with certainty. For this selection function one finds $p = (.6)(.2)+(.3)(.4)+(.1)(.9) = .33$, so that $r = f/(1-p) = 2$.

By choosing m, the number of segments, sufficiently large, one can model any arbitrary selection function to any desired degree of approximation. As an example, consider a discrete approximation to the main diagonal of the unit square,

(15)    $B_k = k/m$,    $A_k = 1/m$,    $k = 1, ..., m$.

For the area under the step-function one finds

(16)    $p = m(m+1)/2m^2$, so that $f/p = r = (m-1)/(m+1) \to 1$ as $m \to \infty$.

This is reasonable because S in eq. (13) converges on the diagonal of the unit square.

It thus emerges that Scargle's argument is perfectly general and not tied to any particular selection function S(z) that may be viewed as unrealistically restrictive. For example, instead of assuming that studies exceeding the cut-off $z_0$ are published with probability 1 as in Fig. 1, some may argue it is more realistic to assume they are published with probability .8. This would necessitate the introduction of a third parameter.

## 4. Conclusions

The fundamental point at issue here is the validity of meta-analyses that relies on combined statistical summaries culled from a literature subject to an unknown degree of publication bias. Essentially, we contest claims that an estimated lower limit to the number of relevant studies denied publication is typically so large as to be unreasonable. We go beyond previous authors who have already noted that Rosenthal&Rubin's FSN derivations were flawed, in claiming that even correctly computed FSNs are typically quite modest for a large parameter region of a bias model which, moreover, is no longer tied to the normal distribution. This meets the burden of proof that publication bias can distort meta-analytic results to the point of solely representing false positives. In view of our generalization to partitions to more than 2 intervals, we believe that our model reasonably approximates the nature of publication bias. By showing that such a large class of plausible bias models can explain any putative meta-analytic conclusion, we have cast reasonable doubt that should be heeded, especially in cases of great social importance – e.g. for clinical medical studies of drug efficacy and safety. Scargle (2000) had already been more explicit than previous authors in questioning the legitimacy, not just of the FSN method, but of meta-analysis in general as a justifiable research tool:

"Statistical combinations of studies from the literature can be trusted to be unbiased only if there is reason to believe that there are essentially no unpublished studies (almost never the case)." (Scargle, 2000, p. 102).

A case in point is the history of the FSN itself, that we reviewed in the Introduction.

References

Darlington, R. (1980) Another peek at the file drawer. Submitted 7/25/1980 to the *Psychological Bulletin*. Rejected 10/27/1981. Personal communication. Letters available from first author upon request.

Elsahoff, J. D. (1978) Commentary. *The Behavioral and Brain Sciences, 1*, 392.

Fisher, R. A. (1948) Combining independent tests of significance. *American Statistician*, 1948, 2, 30.

Iyengar, S.&Greenhouse, J.B. (1988) Selection models and the file-drawer problem. *Statistical Science, 3*, 109-135.

Iyengar, S.&Greenhouse, J.B. (1988) . Rejoinder. *Statistical Science, 3*, 133-135.

Mahoney, M. J. (1977) Publication prejudices: An experimental study of confirmatory bias in the peer review system. *Cognitive Therapy and Research*, 1977, 1, 161-175.

Rosenthal, R. (1979) The "file drawer problem" and tolerance for null results. *Psychological Bulletin*, 86, 638-641.

Rosenthal, R.& Rubin, D. B. (1978) Interpersonal expectancy effects: the first 345 studies. *The Behavioral and Brain Sciences, 1*, 377-416.

Rosenthal, R.& Rubin, D. B. (1988) Comment: Assumptions and procedures in the file-drawer problem. *Statistical Science* (1988) 3, 120-125.

Scargle, J. D. (2000) Publication bias: The "File Drawer" problem in scientific inference. *Journal of Scientific Exploration, 14*, 91-106.

Sutton, A.J, Duval, S.J, Tweedle, R.L., Abrams, K.R., & Jones, D.R. (2000) Empirical assessment of publication bias in meta-analysis. *British Medical Journal*, 320, 1574-

1577.

Thomas, H. (1985) On the "File Drawer" problem. Submitted 1/14/1985 to the *Psychological Bulletin.* Rejected 3/8/1988. Personal communication. Letters available from first author upon request.

Veven, J.L & Woods, C.M (2005) Publication bias in research synthesis: sensitivity analysis using a priori weight functions. Psychological Methods, 10(4), 428-443.

Wallis, W. A. Compounding probabilities from independent significance tests. *Econometrica*, 1942, 10, 229-248.

**Table 1** 14

r := f/p  for selected values of $\alpha$ and $\beta$

| $\beta$ | .01 | .05 | .10 | .15 | .20 | .25 | .30 | .35 | .40 | .45 |
|---|---|---|---|---|---|---|---|---|---|---|
| 1-$\beta$ | .99 | .95 | .90 | .85 | .80 | .75 | .70 | .65 | .60 | .55 |
| $\alpha$  1-$\alpha$ | | | | | | | | | | |
| .01 .99 | 49.25 | 15.81 | 8.17 | 5.31 | 3.81 | 2.88 | 2.26 | 1.81 | 1.46 | 1.20 |
| .02 .98 | 32.56 | 13.49 | 7.47 | 4.99 | 3.63 | 2.77 | 2.18 | 1.75 | 1.43 | 1.17 |
| .03 .97 | 24.19 | 11.74 | 6.87 | 4.70 | 3.46 | 2.67 | 2.12 | 1.71 | 1.39 | 1.14 |
| .04 .96 | 19.16 | 10.36 | 6.35 | 4.43 | 3.31 | 2.57 | 2.05 | 1.66 | 1.36 | 1.12 |
| .05 .95 | 15.81 | 9.26 | 5.90 | 4.19 | 3.17 | 2.48 | 1.99 | 1.61 | 1.33 | 1.09 |
| .10 .90 | 8.17 | 5.90 | 4.26 | 3.26 | 2.57 | 2.08 | 1.70 | 1.41 | 1.17 | .98 |
| .15 .85 | 5.31 | 4.19 | 3.26 | 2.60 | 2.13 | 1.76 | 1.47 | 1.23 | 1.04 | .88 |
| .20 .80 | 3.81 | 3.17 | 2.57 | 2.13 | 1.78 | 1.50 | 1.27 | 1.08 | .92 | .79 |
| .25 .75 | 2.88 | 2.48 | 2.08 | 1.76 | 1.50 | 1.29 | 1.11 | .95 | .82 | .70 |
| .30 .70 | 2.26 | 1.99 | 1.70 | 1.47 | 1.27 | 1.11 | .96 | .83 | .72 | .63 |
| .35 .65 | 1.81 | 1.61 | 1.41 | 1.23 | 1.08 | .95 | .83 | .73 | .64 | .56 |
| .40 .60 | 1.46 | 1.33 | 1.17 | 1.04 | .92 | .82 | .72 | .64 | .56 | .49 |
| .45 .65 | 1.20 | 1.09 | .98 | .88 | .79 | .70 | .63 | .56 | .49 | .43 |
| .50 .50 | .98 | .90 | .82 | .74 | .67 | .60 | .54 | .48 | .43 | .38 |

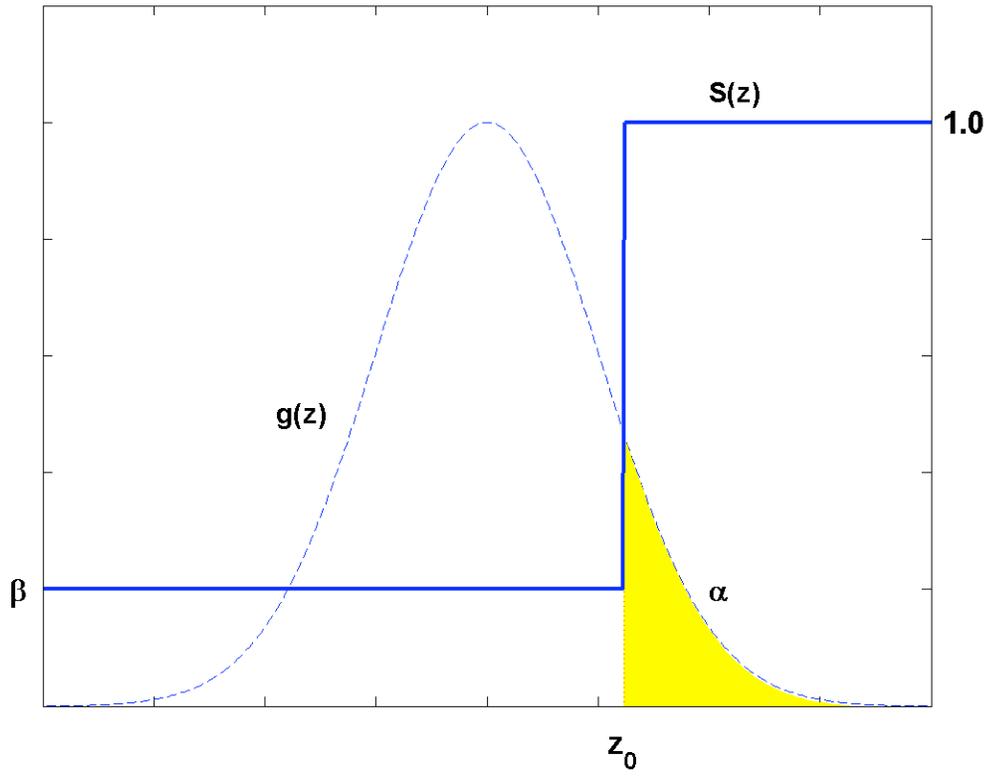

Figure 1

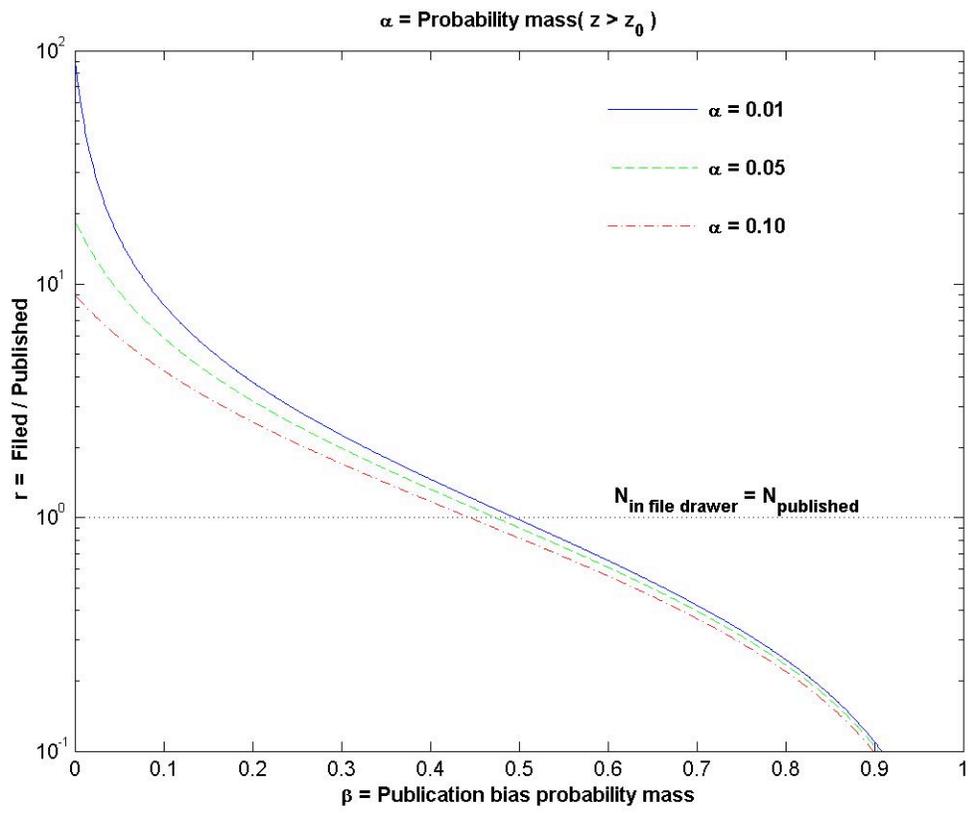

Figure 2

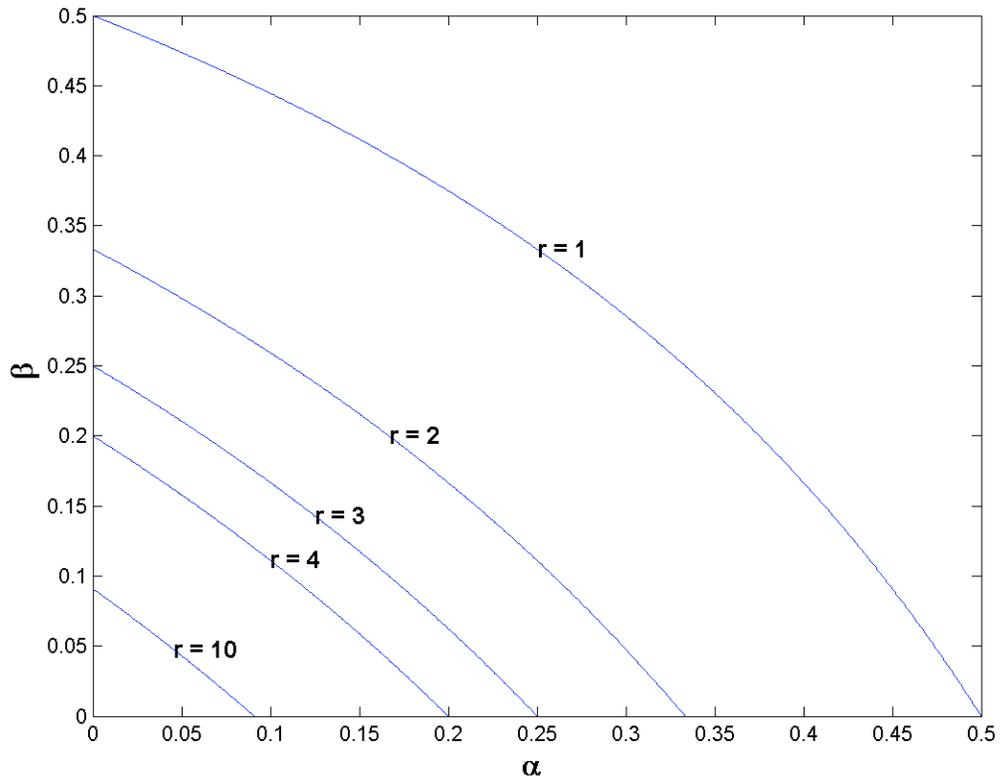

Figure 3

Publication Bias and File-drawer Problem18Figure Captions

Figure 1:

Solid line: Scargle's step-function selection function S(z), representing a publication bias that allows all papers to be published above some threshold, but only a fraction β to be published below. Dashed line: a nominal distribution g(z) is superimposed for illustration: the overall publication probability as a function of z is the product of the two curves, and the shaded area above $z_0$ is the parameter α that we use as a surrogate for $z_0$.

Figure 2:

Plot of ln r, the ratio of unpublished to published papers as a function of the probability mass β describing publication bias, for several common "p-values" α.

Figure 3:

The relation between α and β for selected values of r.

Acknowledgement: We are grateful to Dr. M. Heene, LMU Muenchen, for valuable technical assistance.